\documentstyle[11pt,paspconf,epsfig]{article}
 
\def\>{$>$}
\def\<{$<$}

\def\simlt{\lower.5ex\hbox{$\; \buildrel < \over \sim \;$}}
\def\simgt{\lower.5ex\hbox{$\; \buildrel > \over \sim \;$}}
\def\ch2{$\chi^{2}$}

\def\beq{\begin{equation}}
\def\eeq{\end{equation}}

\def\Cp{C_{\rm peak}}

\newbox\grsign \setbox\grsign=\hbox{$>$} \newdimen\grdimen \grdimen=\ht\grsign
\newbox\simlessbox \newbox\simgreatbox \newbox\simpropbox
\setbox\simgreatbox=\hbox{\raise.5ex\hbox{$>$}\llap
     {\lower.5ex\hbox{$\sim$}}}\ht1=\grdimen\dp1=0pt
\setbox\simlessbox=\hbox{\raise.5ex\hbox{$<$}\llap
     {\lower.5ex\hbox{$\sim$}}}\ht2=\grdimen\dp2=0pt
\setbox\simpropbox=\hbox{\raise.5ex\hbox{$\propto$}\llap
     {\lower.5ex\hbox{$\sim$}}}\ht2=\grdimen\dp2=0pt
\def\simgt{\mathrel{\copy\simgreatbox}}
\def\simlt{\mathrel{\copy\simlessbox}}

\begin{document}

\title{Power Density Spectra of Gamma-Ray Bursts}
 
\author{Andrei M. Beloborodov\altaffilmark{1}}
\affil{Stockholm Observatory, SE-133 36 Saltsj\"obaden, Sweden}

\altaffiltext{1}{Also at the Astro-Space Center of Lebedev Physical Institute,
Profsoyuznaya 84/32, 117810 Moscow, Russia}

\begin{abstract}
Power density spectra (PDSs) of long gamma-ray bursts 
provide useful information on GRBs, indicating their self-similar temporal 
structure.  The best power-law PDSs are displayed by the longest 
bursts ($T_{90}>100$ s) in which the range of self-similar time scales covers 
more than  2 decades. Shorter bursts have apparent
PDS slopes more strongly affected by statistical fluctuations.
The underlying power law can then be reproduced with high accuracy 
by averaging the PDSs for a large sample of bursts. 
This power-law has a slope $\alpha\approx-5/3$ and a sharp break at 
$\sim 1$~Hz.

The power-law PDS provides a new sensitive tool for studies of gamma-ray 
bursts.
In particular, we calculate the PDSs of bright bursts in separate energy 
channels.
The PDS flattens in the hard channel ($h\nu>300$ keV) and steepens in the 
soft channel ($h\nu<50$ keV), while the PDS of bolometric light curves 
approximately follows the $-5/3$ law. 

We then study dim bursts and compare them to the bright
ones. We find a strong correlation between the burst brightness and the PDS
slope. This correlation shows that the bursts are far from being standard 
candles and dim bursts should be {\it intrinsically} weak. 
The time dilation of dim bursts is probably related to physical 
processes occurring in the burst rather than to a cosmological redshift.

Finally, we test the internal shock model against the observed PDS. We 
demonstrate how the model can reproduce the $-5/3$ power law. 

\end{abstract}

\keywords{gamma rays: bursts}

\section{Introduction}

The light curves of gamma-ray bursts (GRBs) typically have many random peaks 
and in spite of extensive statistical studies 
%(e.g., Nemiroff et al. 1994; Norris et al. 1996; Stern 1996), 
the temporal behavior of GRBs remains a puzzle.
Contrary to the complicated diverse behavior in the time domain, long GRBs show 
a 
simple behavior in the Fourier domain (Beloborodov, Stern, \& Svensson 1998). 
Their PDS is a power-law of index $\alpha\approx -5/3$ (with a break at 
$\sim 1$~Hz) plus standard (exponentially distributed) statistical fluctuations 
superimposed onto the power law. The $-5/3$ slope and the 1 Hz break 
characterize the process randomly generating the diverse light curves of GRBs.
Intriguingly, the PDS slope coincides with the Kolmogorov law. 

The power-law behavior is seen in individual bursts (we illustrate
this in Section 3) and may provide a clue to the nature of GRBs. On the other 
hand,
this behavior can be used as a tool in the studies of GRBs.
%In \S 2, we plot PDSs of the four brightest bursts with $T_{90}>100$ s. 
%%The $-5/3$ power law is observed individually in each of these bursts.
%Then, in \S 3, we study the full sample of 527 bursts in channels II+III 
%and calculate their average PDS. 
In particular, one can study the PDS in separate energy channels
(see Section 6). We quantify the difference of the temporal structure
between the channels in terms of the PDS slope and compare the results with 
previous
studies of the autocorrelation function (ACF) in the separate channels. 
In Section 7, we address dim GRBs and compare them to the 
bright ones.  Then, in Section 8, we simulate the observed PDS with the internal 
shock model. The results are discussed in Section 9.
% The details of data analysis are given in Appendix.

%##############################################################################
\section{Data Analysis}

\subsection{The Sample}

In our analysis we use GRB light curves with 64 ms resolution obtained by 
BATSE in the four LAD energy channels, 
I--IV: (I) $20-50$ keV, (II) $50-100$ keV, (III) $100-300$ keV,
and (IV) $h\nu>300$ keV. The background is subtracted in each channel using 
linear fits to the 1024 ms data.

Long bursts are of particular 
interest since their internal temporal structure can
be studied by spectral analysis over a larger range of time scales. We choose 
bursts with durations $T_{90}>20$ s where $T_{90}$ is the time it takes to 
accumulate from 5\% to 95\% of the total fluence of a burst summed over 
all the four channels, I+II+III+IV. 
Hereafter, we measure the brightness of a burst by its peak count rate, $\Cp$,
 in channels II+III.
We analyze bursts with $\Cp>100$ counts per time-bin, $\Delta t=0.064$ s. 
This condition coupled with the duration condition, $T_{90}>20$~s, gives a 
sample of 559 GRBs. The sample still contains GRBs with low fluence, which 
are not good for Fourier analysis. We exclude bursts with fluences 
$\Phi<32\Delta t\Cp$. The resulting sample contains 527 bursts.

%###########################################################################

\subsection{PDS Calculation}

We calculate the Fourier transform, $C_f$, of each light curve, $C(t)$, 
using the standard 
Fast Fourier Transform method. The power density spectrum, $P_f$, is given by 
$P_f=(C_fC_f^*+C_{-f}C_{-f}^*)/2=C_fC_f^*$ since $C(t)$ is real.
The Fourier transform is calculated on a standard grid with a time bin 
$\Delta t=64$ ms and a total number of bins $N_{\rm time}=2^{14}$, which 
corresponds to a total time $T\approx 1048$ s since the trigger time. 
The light curve of each burst is considered in its
individual time window $(t_1,t_2)$.
% (see Section 2.4). 
In the calculations of the PDS,
we extend the time interval to $(0,T)$ by adding zeros at $(0,t_1)$ and 
$(t_2,T)$. 
%The adding zeros introduces random small-scale fluctuations in the 
%PDS. We, however, study the PDSs averaged
%over adjacent frequencies and/or over a sample of GRBs. Then the random 
%fluctuations in $P_f$ associated with a specific choice of the grid disappear 
%and do not affect the results. 
See Beloborodov, Stern, \& Svensson (1999) for details of the data analysis.

%\subsection{The Poisson Level}

Poisson noise in the measured count rate affects the light curve on short
time scales, and, correspondingly, affects the PDS at high frequencies.
The Poisson noise has a flat spectrum, $P_f\propto f^0$, introducing 
the ``Poisson level'', $P_0$, in a PDS. This level equals the burst total 
fluence 
including the background in the considered time window. The power spectrum above 
this level displays the intrinsic variability of the signal (but see Section 5 
for
the time-window effects). We calculate the 
individual Poisson level for each burst and subtract it from the burst PDS.

%\subsection{The Time Window}

%One would like to see the whole burst in the window. However, 
%(i) it is difficult to determine exactly the end of a burst because the burst 
%can always have a weak ``tail'' hidden in the background;
%(ii) the window should not be very large because inclusion of long weak tails 
%leads to an increase of the background fluence, $\Phi_b$, without a substantial 
%increase in the signal fluence, $\Phi$. The resulting low ratio 
%$\Phi/\Phi_b$ implies a high Poisson level, which 
%makes the quality of the PDS worse.
%
%To reduce the Poisson level, we cut off the light curves at time $t_{\rm max}$ 
%defined so that the signal count rate $C(t)$ does not exceed 
%$\varepsilon C_{\rm peak}$ at $t>t_{\rm max}$ and 
%$C(t)=\varepsilon C_{\rm peak}$ at $t=t_{\rm max}$. 
%Keeping in mind bursts with a weak beginning, we define the starting window 
%time, $t_{\rm min}$, so that $C(t)<\varepsilon C_{\rm peak}$ at 
%$t<t_{\rm min}$ and $C(t)=\varepsilon C_{\rm peak}$ at $t=t_{\rm min}$. 
%In our analysis $\varepsilon=0.05$ is chosen. We checked that varying
%$\varepsilon$ does not significantly affect the results unless 
%$\varepsilon>0.1$. Note that, for dim bursts, Poisson fluctuations of the 
%background (which are imprinted on $C(t)$ even after subtraction of the average
%background level) may exceed $\varepsilon\Cp$. Then the window is determined by
%the background fluctuations rather than by the signal. 

%##############################################################################

\section{Individual Bright Bursts}

The brightest and longest bursts are the best ones for Fourier analysis.
In this section we study the four brightest bursts with $T_{90}>100$ s.
They have trigger numbers \# 2156 (GRB 930201), 2856 (GRB 940302), 3227 
(GRB 941008), and 6472 (GRB 971110).

We take the light curves, $C(t)$, summed over channels II and III, in which the 
signal is strongest. To simplify the comparison of different bursts,
we take peak-normalized light curves. Their Fourier transform, $C_f$,
is therefore normalized by $\Cp$, and the PDS, $P_f=C_fC_f^*$, is normalized 
by $\Cp^2$.

The four peak-normalized light curves and their PDSs are shown in Figure~1 
(we smooth the PDSs on the scale $\Delta \log f = 0.04$ before plotting).
The light curves are very different while the PDSs are similar. They can be 
described as a single power law, $\log P_f=A+\alpha\log f$ ($A\approx 1$ and 
$\alpha\approx -1.5$) with super-imposed fluctuations $\Delta P_f/P_f\sim 1$. 
In spite of the large $\Delta P_f$, the power-law behavior can be seen in 
each burst due to the power-law extending over more than two decades in 
frequency. 

%%%%%%%%%%%%%%%%%%%%%%%%%%%%%%%%%%%%%%%%%%%%%%%%%%%%%%%%%%%%%%
\begin{figure*}
\begin{center}
\epsfxsize=13.5cm
\epsfysize=13.5cm
\epsfbox{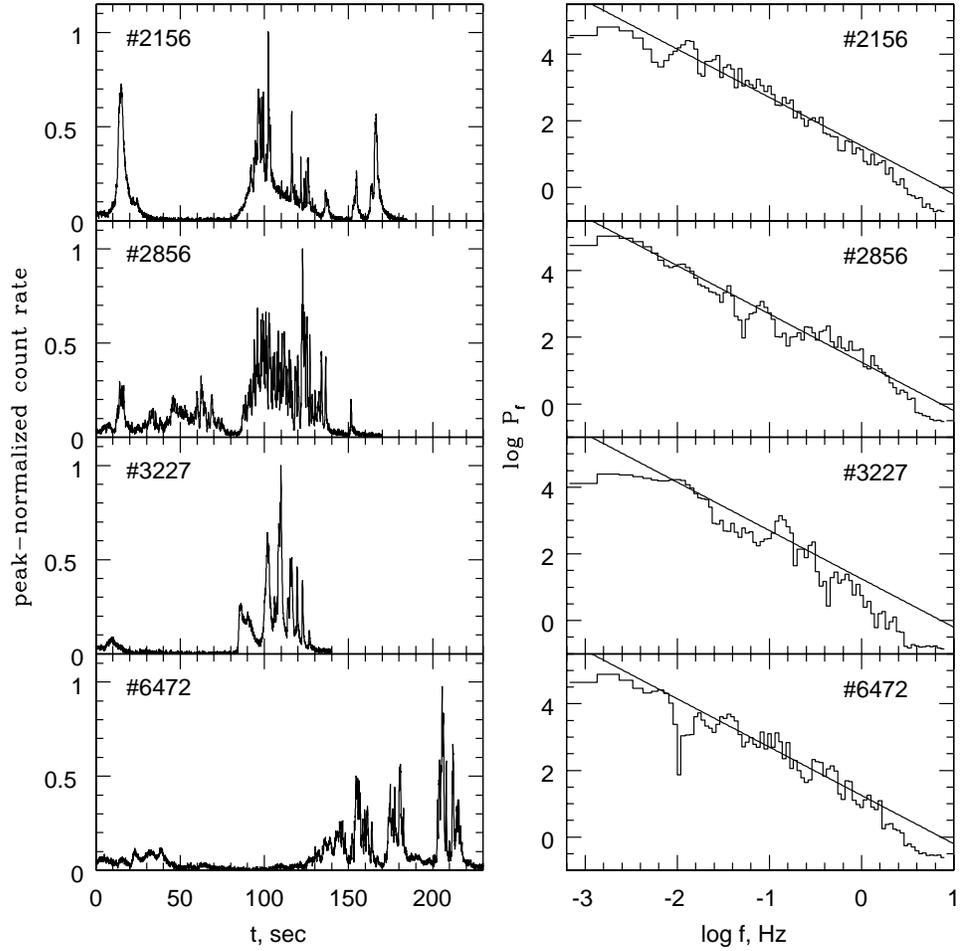}
\end{center}
\caption{The light curves (in channels II+III) and their PDSs for the four 
brightest bursts in the sample with $T_{90}>100$ s. The light curves are taken 
peak-normalized and, correspondingly, the PDSs are normalized by $\Cp^2$. 
%The PDSs are smoothed on the scale $\Delta \log f = 0.04$. 
%The horizontal lines shows the individual Poisson levels (divided by $\Cp^2$).
Since the bursts are very bright, their Poisson levels (normalized by $\Cp^2$)
are very low, $P_0\approx 10^{-4}$. 
The straight lines show the fit to the average of the 4 PDSs, 
$\log P_f=1.25-1.45\log f$ (see Fig.~2).
%Subtraction of $P_f^0$ hardly changes $P_f$.
}
\end{figure*}
%%%%%%%%%%%%%%%%%%%%%%%%%%%%%%%%%%%%%%%%%%%%%%%%%%%%%%%%%%%%%%

The presence of an underlying power law is an interesting feature of the GRB
temporal behavior. A possible way 
to subtract it from the noisy individual PDSs is to take
the average PDS over a sample of long GRBs. Then the 
fluctuations affecting each individual PDS tend to 
cancel each other and one can see the power law. 
The average of the four PDSs is plotted in Figure~2.
%shows the average PDS for the four bursts. 
One can see that the amplitude of irregularities is reduced after the 
averaging and one can measure the slope of the resulting PDS. The best 
power-law fit in the range $-2.0<\log f<0.2$ gives 
$\log P_f = 1.25 - 1.45 \log f$. 

%%%%%%%%%%%%%%%%%%%%%%%%%%%%%%%%%%%%%%%%%%%%%%%%%%%%%%%%%%%%%%
%\medskip
\begin{figure}[t]
\centerline{\epsfxsize=9.0cm\epsfysize=9.0cm {\epsfbox{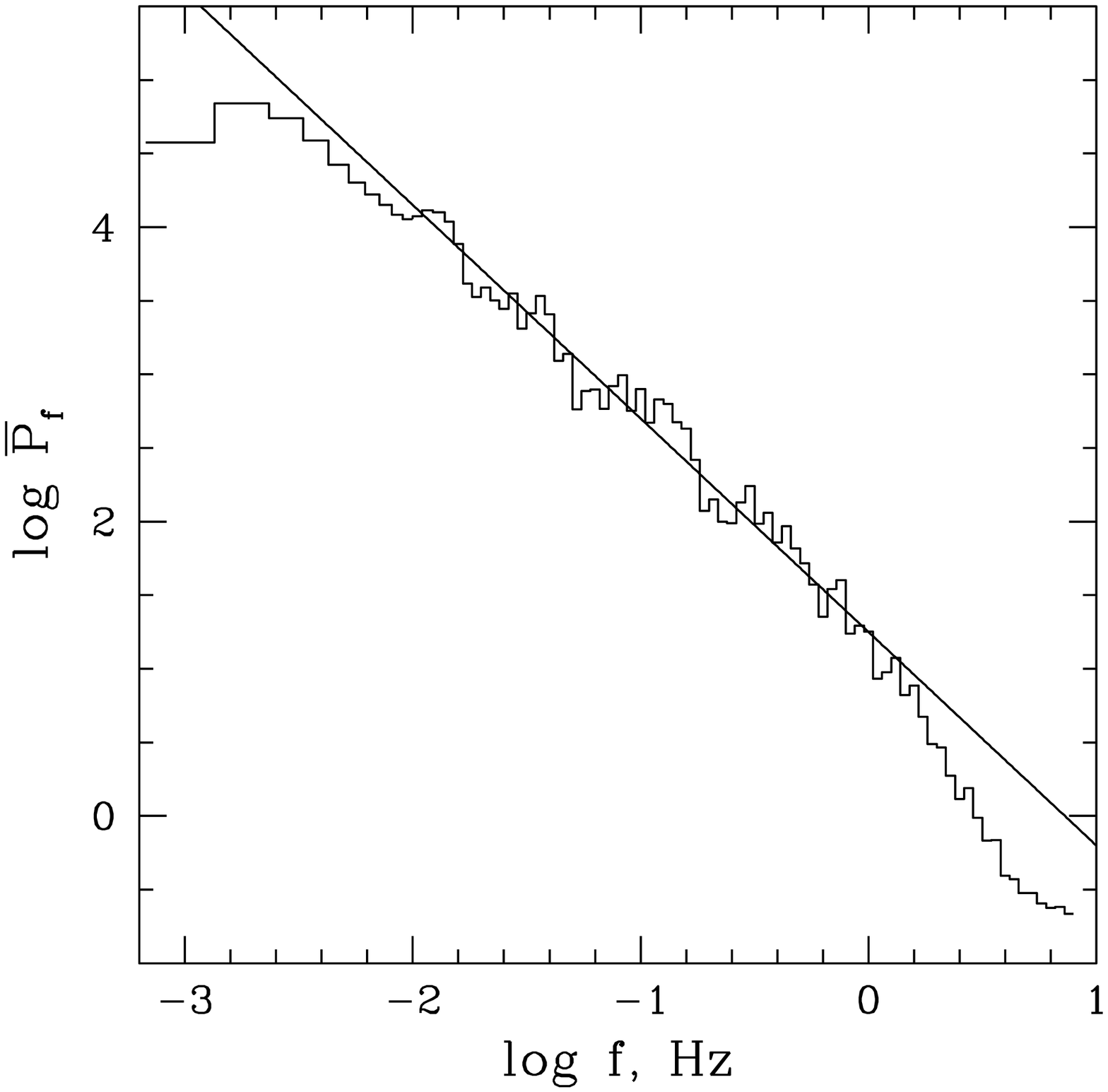}} }
\caption{ The average of the 4 PDSs shown in Figure~1. The line
shows the best power-law fit, $\log P_f=1.25-1.45\log f$, in the 
range $-2.0 < \log f < 0.2$.
\label{fig2}
}
\end{figure}
%\bigskip
%%%%%%%%%%%%%%%%%%%%%%%%%%%%%%%%%%%%%%%%%%%%%%%%%%%%%%%%%%%%%%

The PDS analysis was previously performed for a number of individual bursts. 
Belli (1992), for example, analyzed GRBs detected by
the Konus experiment onboard the Soviet probes Venera 11 and Venera 12, and
Giblin et al. (1998) studied individual BATSE bursts.
The power-law PDS can be observed in their longest complex bursts as well. 

When turning to GRBs with relatively modest durations, $T_{90}\sim 20 -100$~s,
one finds it more difficult to see the power law in individual bursts.
The statistical fluctuations, $\Delta P_f/P_f\sim 1$, significantly
affect the apparent PDS slope derived from the available range of frequencies, 
which is limited by $\sim T_{90}^{-1}$. However, having a large number of 
bursts allows one to study the PDS by averaging over the sample. 

%%%%%%%%%%%%%%%%%%%%%%%%%%%%%%%%%%%%%%%%%%%%%%%%%%%%%%%%%%%%%%
\smallskip
\begin{figure}
\centerline{\epsfxsize=9.0cm\epsfysize=8cm {\epsfbox{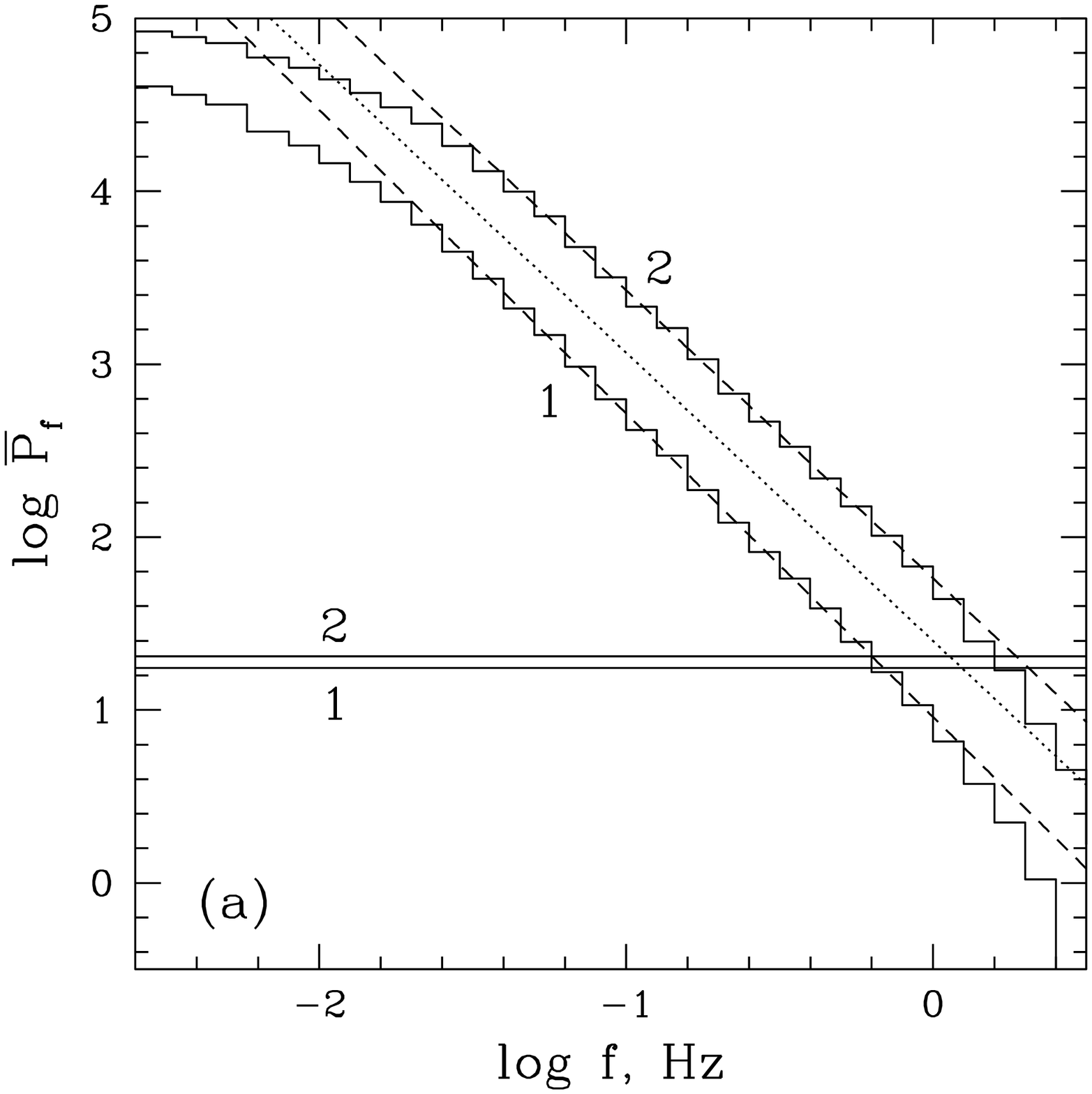}} }
\centerline{\epsfxsize=9.0cm\epsfysize=8cm {\epsfbox{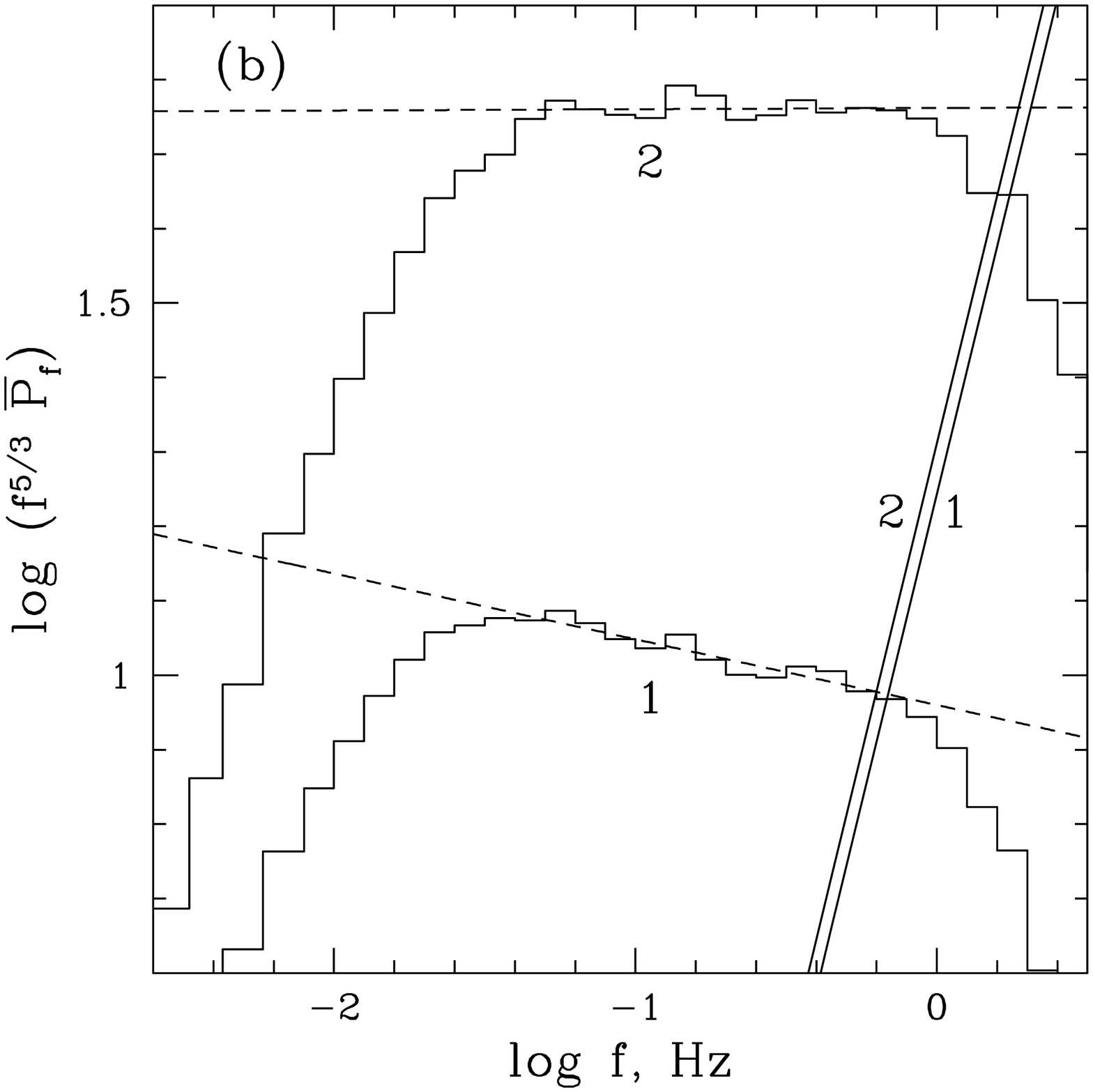}} }
%\bigskip
\caption{ The average PDS (in channels II+III) for the full sample of 
527 GRBs. (a) -- The histograms marked as 1 and 2 correspond to the 
peak-normalization and $\sqrt{\Phi}$-normalization, respectively.
The horizontal lines show the corresponding normalized Poisson levels averaged
over the sample. {\it Dashed lines} are the best power-law fits in the range
$-1.4<\log f<-0.1$: $\log\bar{P}_f=0.96-1.75\log f$ in the peak-normalized
case and $\log\bar{P}_f=1.76-1.67\log f$ in the $\sqrt{\Phi}$-normalized 
case. {\it Dotted line} shows the $-5/3$ slope.
(b) -- Same as (a) except that now we plot $\bar{P}_f$ multiplied by $f^{5/3}$.
%({\it dotted histogram} -- before subtraction of the Poisson level, 
%{\it solid histogram} -- after subtraction). The solid line shows the 
%average normalized Poisson level for the sample, $\bar{P}_0=17.0$.
\label{fig3}}
\end{figure}
%\bigskip
%%%%%%%%%%%%%%%%%%%%%%%%%%%%%%%%%%%%%%%%%%%%%%%%%%%%%%%%%%%%%%

%##############################################################################

\section{The Average PDS}

In most of the GRB models (e.g., in the internal shock model) different bursts 
are 
produced by one physical mechanism of stochastic nature, i.e., an individual 
burst
is a random realization of the same standard process. Stern \& Svensson (1996) 
suggested an empirical pulse-avalanche model of this type. They showed that 
near-critical avalanches reproduce the statistical characteristics of GRBs 
as well as their extreme diversity. With such an approach, individual GRBs are 
like pieces of a ``puzzle'', and the features of the standard engine can be 
probed 
with statistical methods applied to a sufficiently large ensemble of GRBs. 

The simplest
statistical characteristic of the PDS is the average PDS, $\bar{P}_f$.
The averaging means that we sum up the PDSs of individual bursts with some
weights (normalization) and divide the result by the number of bursts in the 
sample. If no normalization is employed, then
the brightest bursts strongly dominate $\bar{P}_f$, 
and their individual fluctuations lead to strong fluctuations in $\bar{P}_f$. 
The PDSs of relatively weak bursts are suppressed
proportionally to $\Cp^{-2}$ and they are lost in the averaging.
A normalization procedure is thus needed to increase the weight of
relatively weak bursts in the sample. We prefer the peak-normalization because: 

(i) The amplitude of the resulting PDS is practically independent of the burst 
brightness (see, e.g., Fig. 1).

(ii) The fluctuations in $\bar{P}_f$ turn out to be minimal and the best 
accuracy 
of the slope is achieved. The distribution of individual $P_f$ around 
$\bar{P}_f$ 
follows a standard exponential law (see Beloborodov et al. 1998), and the 
amplitude
of fluctuations in $\bar{P}_f$ is given by a simple formula, 
$\Delta \bar{P}_f/\bar{P}_f \sim N^{-1/2}$, where $N$ is the number of bursts 
in the sample.

The averaging procedure is the following.
(1) For each burst we first determine its individual Poisson level.
% which equals the total fluence of signal + background in the considered 
% time window.
(2) We determine the peak, $\Cp$, of the light curve, $C(t)$, with the 
background
    subtracted.
(3) We normalize the light curve to its peak.
(4) We calculate the PDS of the normalized light curve. 
(5) We normalize the Poisson level by $\Cp^2$ and 
    subtract it from the PDS.
(6) We sum up the resulting PDSs over the sample and divide by the number of 
bursts. The resulting  average PDS, $\bar{P}_f$, for the 527 peak-normalized 
light curves in channels (II+III) is shown in Figure~3. 

For comparison, we also plot the average PDS with the 
$\sqrt{\Phi}$-normalization, where $\Phi$ is the burst fluence excluding the
background. Each light curve is then normalized by $\sqrt\Phi$, and its PDS (and
its Poisson level) is normalized by $\Phi$. This normalization gives different 
weights (the weights of dim bursts are reduced as compared to the
peak-normalization). The resulting average PDS has a different amplitude.
Nevertheless, $\bar{P}_f$ again follows a power-law with approximately the
same slope. This provides evidence that the self-similar behavior with 
$\alpha\approx -5/3$ is an intrinsic property of GRBs, rather than an 
artefact of the averaging procedure. This interpretation
is also supported by the fact that we observe the power-law in the longest
individual bursts. When comparing individual PDSs, $P_f$, with the 
peak-normalized
$\bar{P}_f$, one sees that $P_f$ is exponentially distributed around 
$\bar{P}_f$,
and the distribution is self-similar with respect to shifts in $f$.

The power-law fit to $\bar{P}_f$ in the range $-1.4 <\log f < -0.1$ gives 
$\alpha=-1.75$ for 
the peak-normalized bursts, and $\alpha=-1.67$ for $\sqrt{\Phi}$-normalized 
bursts. Note that the slope in the 
peak-normalized case is different from $-1.67$ reported in Beloborodov et al. 
(1998). This change is caused by that we have extended the sample to smaller 
brightnesses (see Section 7).

The deviation from the power-law at the low-frequency end is due to the 
finite duration of bursts.
% (the average $T_{90}$ is about 80 s for our sample).
%
%Poisson noise in the light curves affects $\bar{P}_f$ at high $f$.
%To see the behavior of $\bar{P}_f$ at high frequencies,
%one must subtract the individual Poisson level in each burst
%before the averaging. This is mathematically equivalent to subtraction of the 
%effective average level (shown in Fig.~3) from the average PDS. 
%%The importance of the Poisson level subtraction is illustrated in 
%%the bottom panel in Figure~3. 
%After the subtraction, a break appears at about 1 Hz. 
At the high frequency end, there is a break at $\sim 1$~Hz.
The break is observed in the brightest GRBs even without 
subtracting the Poisson level (see Beloborodov et al. 1998 and the top
panel in Fig.~7). Note that the break position is the same for the
peak-normalized and $\sqrt{\Phi}$-normalized bursts. It stays the same 
in the separate energy channels (see Fig.~5) and does not depend on the Poisson 
level. One may also see the break in individual long bursts (Figs.~1 and 2).
The break is far too sharp to be explained as an artefact of 
the 64 ms time binning, which suppresses the PDS by a factor of
$[\sin(\pi f\Delta t)/(\pi f \Delta t)]^2$ where $\Delta t=64$ ms is the time 
bin (cf. van der Klis 1989).

%#############################################################################

\section{The Effects of Finite Signal Duration} 

Fourier analysis was designed for physical problems dealing with 
linear differential equations. For example, it is usually applied to small 
perturbations above a given background solution. The Fourier power spectrum
is also commonly used in the temporal studies of long signals or noises,
e.g., in persistent astrophysical sources. By contrast, GRBs
are strongly non-linear and short signals. The typical number of BATSE 
time-bins ($\Delta t=64$ ms) in a long GRB is a few $\times 10^3$.
%Could one hope to measure an intrinsic power law in GRB power spectra? 
%Or 
Do the effects of finite duration (i.e., time-window effects) 
strongly affect the measured PDS? 

The issue is illustrated in Figure~4. We prepared an artificial long 
signal with $\bar{P}_f\propto f^{\alpha}$ and exponentially distributed $P_f$, 
i.e., the probability to detect $P_f$ at a given $f$ is proportional to 
$\exp(-P_f/\bar{P}_f$). The phase structure was taken to be Gaussian (random).
The signal duration is $T_0=2^{17}\Delta t\approx 8400$~s. Then, we cut the long 
signal into 64 pieces of equal length $t_0=2^{11}\Delta t\approx 130$ s.
We thus get 64 random short signals, each is a random realization/fragment of 
the same stationary process characterized by the index $\alpha$.
Analogously to our analysis of real GRBs, we normalize each signal to its peak 
and add zeros up to $T=2^{14}\Delta t$ (our standard grid). Then we calculate 
the average PDS, $\bar{P}_f$, for the 64 artificial signals. The result is 
shown in Figure~4 for the three cases: $\alpha=0$ (Poisson), $\alpha=-1$ 
(flicker), and $\alpha=-5/3$ (Kolmogorov). 

%%%%%%%%%%%%%%%%%%%%%%%%%%%%%%%%%%%%%%%%%%%%%%%%%%%%%%%%%%%%%%
\smallskip
\begin{figure}[t]
\centerline{\epsfxsize=12cm\epsfysize=12cm {\epsfbox{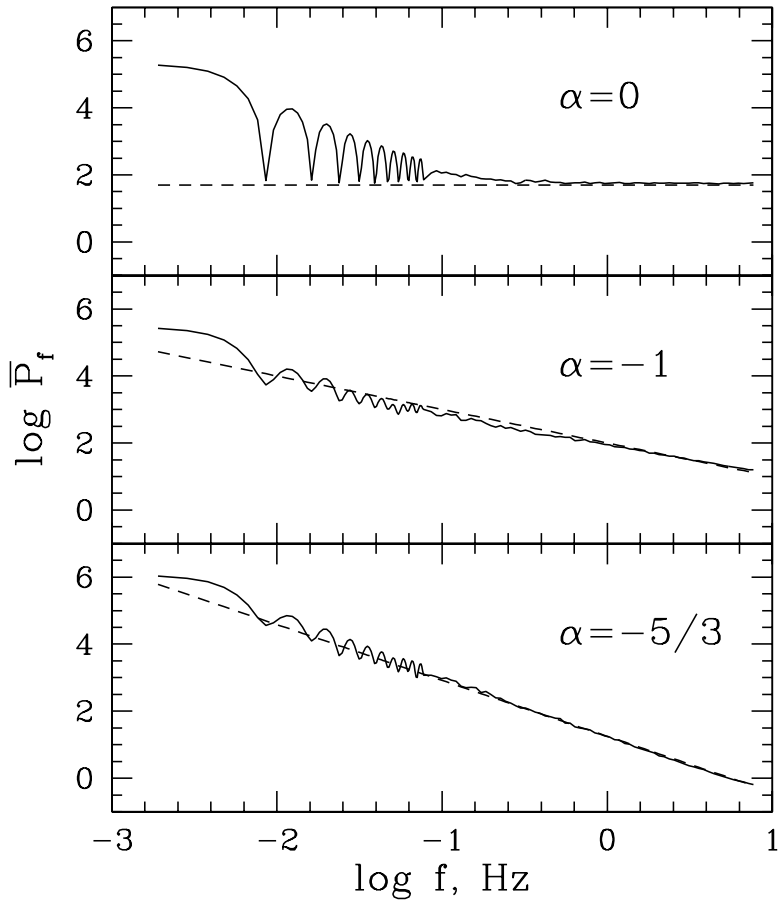}} }
%\bigskip
\caption{ The average PDS, $\bar{P}_f$, for the 64 fragments of noise
with a power-law PDS of index $\alpha$ (see text for details). {\it Dotted 
lines} show the original PDS slope. 
The deviation of $\bar{P}_f$ from this slope is due to the time-window effects.
\label{fig4}}
\end{figure}
%\bigskip
%%%%%%%%%%%%%%%%%%%%%%%%%%%%%%%%%%%%%%%%%%%%%%%%%%%%%%%%%%%%%%

One can see that the time-window effects are strong in the case $\alpha=0$. 
%This is an obvious consequence of that variations on 
%large time scales, $t$, are suppressed as $t^{-1/2}$ for a Poisson signal,
%i.e., 
The Poisson signal is roughly constant on large time scales. As a result, at 
modest 
$f$, $\bar{P}_f$ is just the power spectrum of the $t_0$-window. 
The PDS shape in the case $\alpha=0$ can easily be calculated analytically. 
The Poisson signal can be modelled as a sum of $N$ elementary narrow pulses
($\delta$-functions), with each pulse having equal probability to appear at 
any moment $t$ within the interval $(0,t_0)$. The Fourier transform of one
random realization $(t_1,...,t_N)$ of the signal is 
$C_f=\Sigma \exp(-i 2\pi ft_k)$, and $P_f=C_fC_f^*$. After averaging over 
many realizations, one gets 
$$
\bar{P}_f=N+\frac{N(N-1)\sin^2(\pi ft_0)}{(\pi f t_0)^2}.
$$
The time-window effects are dominant at $f<\sqrt{N}/(\pi t_0)$. 
At $f>\sqrt{N}/(\pi t_0)$,
$\bar{P}_f$ approaches the Poisson level, $N$.

By contrast, in the case $\alpha=-5/3$, we have large-amplitude variations
in the signal on all time scales. The average PDS of the 64 short signals then
reproduces well the intrinsic PDS slope throughout the whole range of 
frequencies,
down to $f\sim t_0^{-1}$ (see bottom panel in Fig.~4). 
Hence, the power law we observe in the average PDS of GRBs 
can be interpreted as that GRBs are random short realizations of a  
standard process which is characterized by the PDS slope $\alpha\approx -5/3$. 
Note, however, that the power spectrum does not provide a complete description
of the signal since the phase structure is not considered. 

%#############################################################################

\section{The PDSs and the ACFs in Channels I, II, III, IV}  

What does the average PDS look like in separate energy channels I, II, III, and 
IV? 
The signal to noise ratio is low in channel I and especially in 
channel IV. The number of bursts for which good PDSs can be obtained in 
channels I and IV is therefore limited to the brightest bursts. 
In order to compare the PDSs in different channels, we choose a sample
of bursts with $\Cp>500$ counts/bin (in channels II+III), which contains 152 
bursts. 

The burst analysis is now performed in each channel separately.
We determine the Poisson level of a burst in each channel, $P_0^i$, and 
find the peak of the light curve, $\Cp^i$, where 
$i=$ I, II, III, IV. Then we perform the peak-normalization: 
$C^i(t)\rightarrow C^i(t)/\Cp^i$ and $P_0^i\rightarrow P_0^i/(\Cp^i)^2$.
%We set the time window in each channel $(t_1^i,t_2^i)$ as described
%in Section 2.4. 

The resulting average PDSs in channels I--IV are shown in Figure~5.
The differences in the slopes are clearly seen.
We fitted the PDSs by power laws, $\log\bar{P}_f=A+\alpha\log f$, in the 
range $-1.6 < \log f < 0$. Channel IV is fitted in the range $-1.3 < \log f < 
-0.1$.
$A$ and $\alpha$ of the best fits are listed in Table~1.

%%%%%%%%%%%%%%%%%%%%%%%%%%%%%%%%%%%%%%%%%%%%%%%%%%%%%%%%%%%%%%
\smallskip
\begin{figure}
\centerline{\epsfxsize=17cm\epsfysize=17cm {\epsfbox{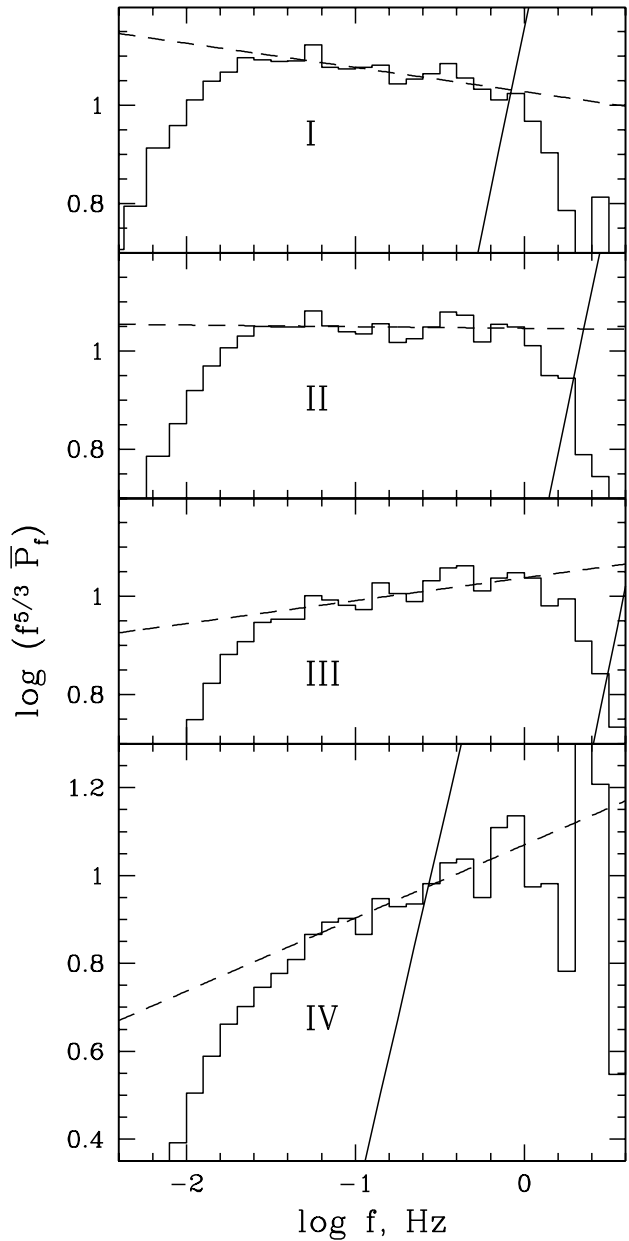}} }
%\bigskip
\caption{ The average PDS for the sample of 152 GRBs in separate
energy channels, I, II, III, and IV. {\it Dashed lines} show the best
power-law fits (see Table~1). {\it Solid lines} show the Poisson level.
\label{fig5}}
\end{figure}
%\bigskip
%%%%%%%%%%%%%%%%%%%%%%%%%%%%%%%%%%%%%%%%%%%%%%%%%%%%%%%%%%%%%%

\medskip
%*************************************************
%\begin{table}
%\caption[ ]{The efficiency of $\gamma$-ray transformation into an $e^\pm$-wind}
\begin{center}
%{\normalsize
\begin{tabular}{ccccc}
            &   {\bf Table~1.}  &         &        \\ [0.4ex]
\hline\hline
            &                   &         &        \\ [0.4ex]
  Channel   &    I    &   II    &   III   &    IV  \\ [0.8ex]
  $\alpha$  & $-1.72$ & $-1.67$ & $-1.60$ & $-1.50$\\ [0.8ex]
  $A$       &   1.03  &  1.05   &   1.06  &   1.07 \\ [0.8ex]
\hline
\end{tabular}
%}
\end{center}
%\end{table}
%*************************************************
\medskip

The observed behavior can be briefly described as follows: The ``red'' power 
($\bar{P}_f$ at low $f$) decreases at high photon energies and increases at low 
photon energies. 
It should be compared with the well-known fact that the pulses in a GRB are more 
narrow in the hard channels (e.g., Norris et al. 1996). The hardness of emission
varies dramatically during a burst and this leads to different temporal 
structure
in different channels. 
E.g., pulses observed in the soft channels may be suppressed in the hard 
channels.
One therefore could expect that the PDS has different slopes in different 
channels.
Note that, typically, most of the GRB energy is released in channels II+III, 
and the average PDS of {\it bolometric} light-curves approximately follows 
the $-5/3$ law.

In principle, the autocorrelation function (ACF) contains the same information
as the PDS, since one is the Fourier transform of the other (the Wiener-Khinchin
theorem). In practice,
the two are not completely equivalent because of the time-window effects and
the presence of a noisy background. On modest time scales, $\simlt 30$~s, the
direct ACF calculation and the calculation via the Fourier transform of the PDS
give the same result to within a few percent (we have calculated the ACF by both
methods). The average ACF, $\bar{A}(\tau)$, for our sample of 152 bright GRBs
is shown in Figure~6 in each of the four channels. The ACF gets narrow at high
energies, in agreement with previous studies (Fenimore et al. 1995), except
that our ACF is $\sim 2$ times wider as compared to that obtained by Fenimore
et al. (1995). The ACF width averaged over the channels is in approximate
agreement with that calculated for the bolometric light curves by Stern \& Svensson
(1996).

%%%%%%%%%%%%%%%%%%%%%%%%%%%%%%%%%%%%%%%%%%%%%%%%%%%%%%%%%%%%%%
%\medskip
\begin{figure}
\centerline{\epsfxsize=8cm\epsfysize=8cm {\epsfbox{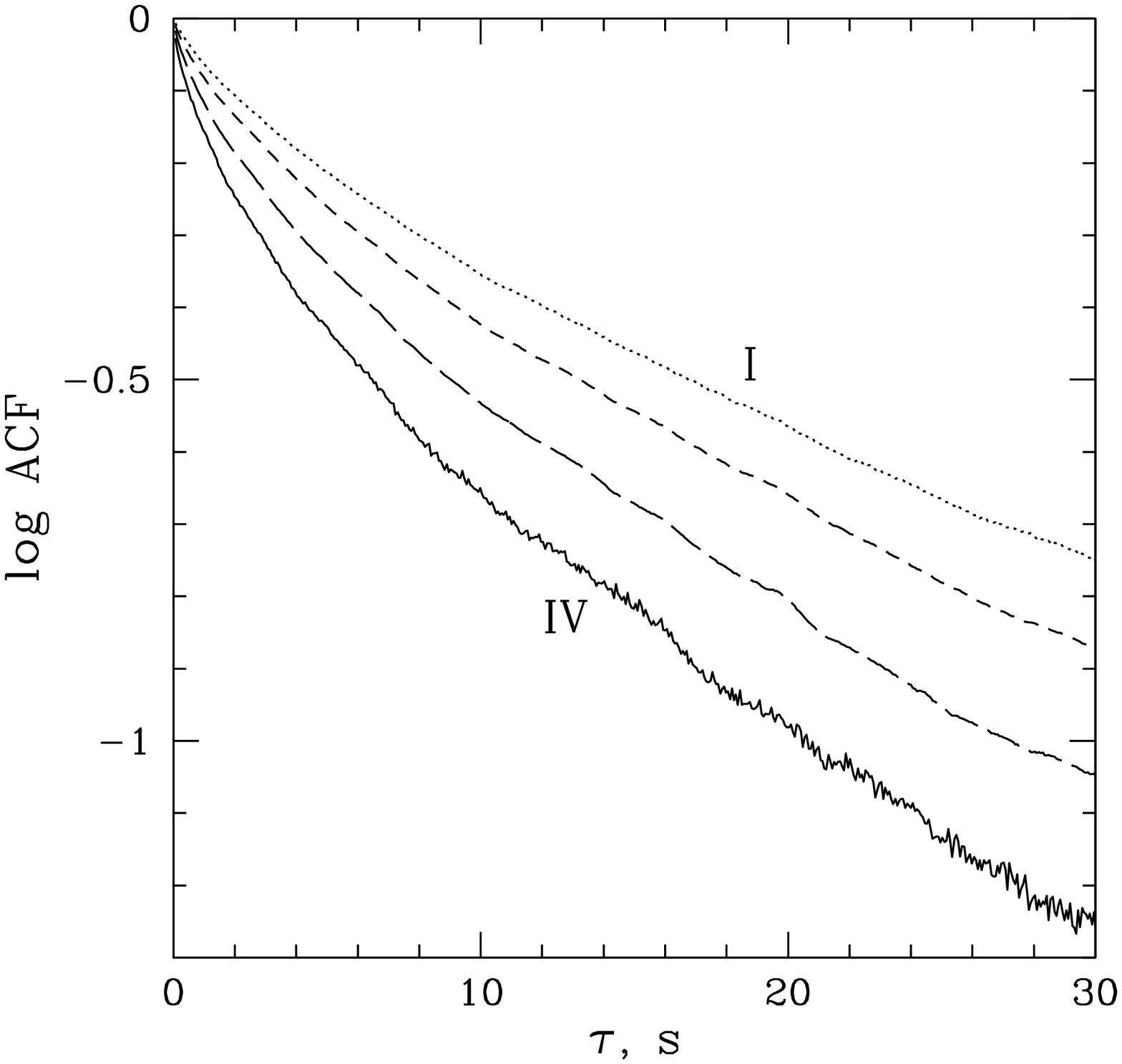}} }
\centerline{\epsfxsize=10cm\epsfysize=11cm {\epsfbox{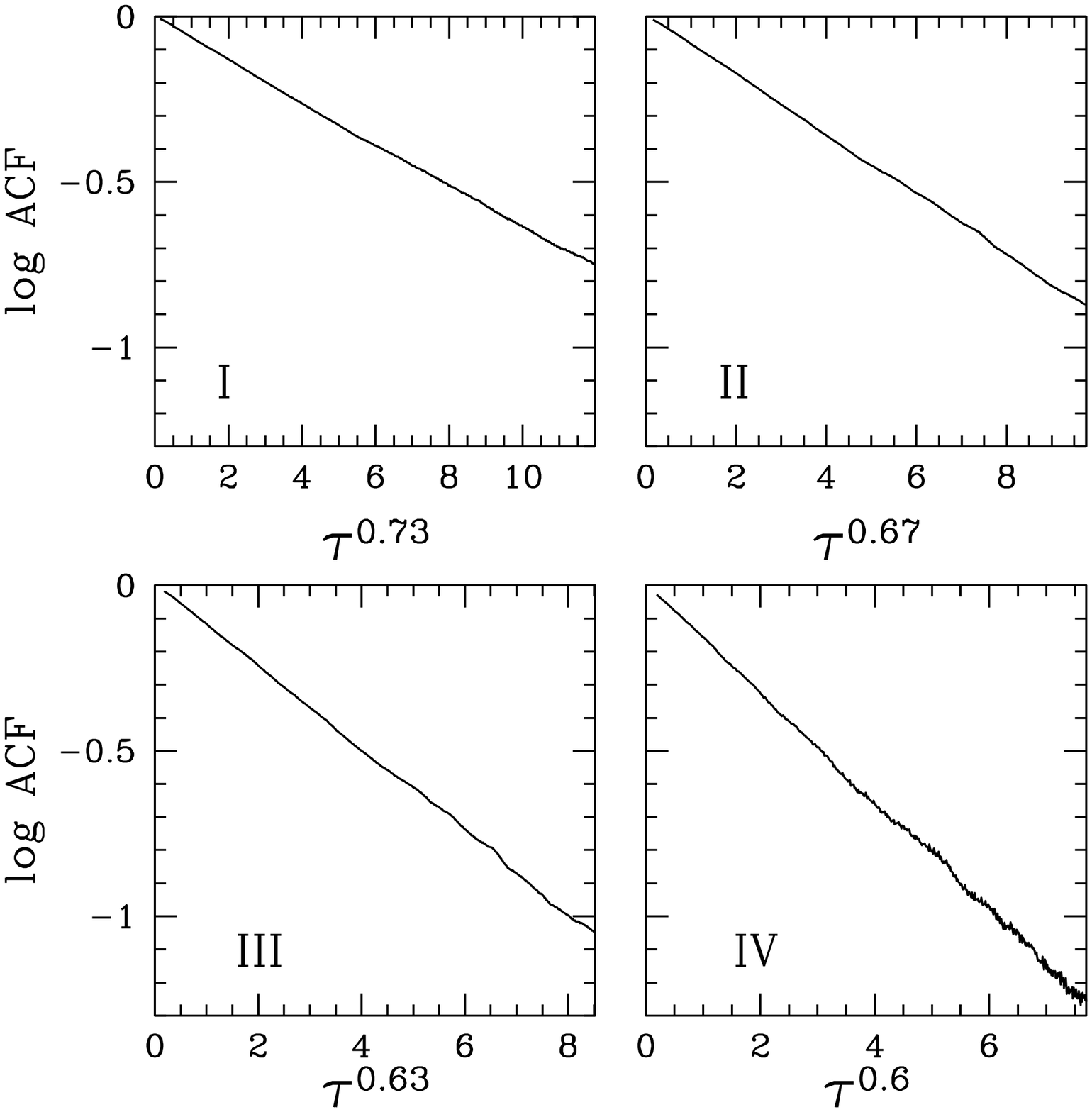}} }
\caption{ The average autocorrelation function for the sample of 152 bright
bursts, in channels I--IV.
%: {\it dotted curve} -- channel I, {\it short-dashed curve} -- channel II,
%{\it long-dashed curve} -- channel III, and {\it solid curve} -- channel IV.
}
\end{figure}
%\bigskip
%%%%%%%%%%%%%%%%%%%%%%%%%%%%%%%%%%%%%%%%%%%%%%%%%%%%%%%%%%%%%%

Note that the average ACF is obtained by summing up the
individual ACFs normalized to unity, i.e., $A(0)=1$. Recalling that the ACF is
the Fourier transform of the PDS, one can see that this normalization is equivalent
to the normalization of the light curve by $\sqrt{P_{\rm tot}}$ where
$P_{\rm tot}=\int [C(t)]^2 dt = \int P_f df$ is the total power. This
normalization is {\it different} from the peak-normalization we use in the
calculations of the average PDS, i.e., we prescribe different weights to
individual bursts when averaging the PDS and the ACF, respectively. Therefore, the average
ACF is {\it not} the Fourier transform of the average PDS shown in Figure~5.
This relation holds when the average PDS is calculated with the
$\sqrt{P_{\rm tot}}$-normalization.

The ACF in each channel is perfectly fitted by the stretched exponential:
$\bar{A}(\tau)=\exp(-[\tau/\tau_0]^\beta)$ (see Figure 6). The parameters
$\tau_0$ and $\beta$ are listed in Table~2. The index $\beta$ is related to
the PDS slope $\alpha$ by the simple relation $\beta\approx -(1+\alpha)$.

\medskip
%*************************************************
%\begin{table}
\begin{center}
%{\normalsize
\begin{tabular}{ccccc}
            &   {\bf Table~2.}  &         &        \\ [0.4ex]
\hline\hline
            &                   &         &        \\ [0.4ex]
  Channel   &    I    &   II    &   III   &    IV  \\ [0.8ex]
  $\beta$   & $0.73$  & $0.67$  & $0.62$  &  $0.6$ \\ [0.8ex]
  $\tau_0$  & $14.0$  & $10.7$  & $7.3$   &  $5.1$ \\ [0.8ex]
\hline
\end{tabular}
%}
\end{center}
%\end{table}
%*************************************************
\medskip

The changing $\beta$ demonstrates that the ACF shape changes
from channel to channel, as it should do since the PDS changes.
As a first approximation, the PDS slope is equal to $-5/3$, and the ACF
index $\beta\approx 2/3=5/3-1$.

The values of $\tau_0$ quantify the ACF width in different channels.
The parameter $\tau_0$ is better defined as compared to measuring
the ACF width at a certain level, e.g., at $e^{-0.5}$ of the maximum level, as
was done in Fenimore et al. (1995). Nevertheless, the scaling of $\tau_0$ with
energy is approximately the same, $\tau_0\propto E^{-0.4}$ where $E$ is the
photon energy of the low energy boundary of the channel, see
Fenimore et al. (1995). It should be stressed, however, that the stretching of
$\tau_0$ is {\it not} related to the stretching of any physical time scale of
intrinsic correlations during the burst. The ACF width, $\tau_0$, is
rather related to the position of the {\it breaks} in the PDS, especially with
the low frequency break, which in turn is determined by the burst duration.
For an infinitely long signal with a power law PDS, $t_0$ would tend to
infinity. This is natural since the power law behaviour implies self-similarity,
i.e., the absence of any preferred time scale. Specific time scales are introduced
only by the breaks in the PDS. We therefore have only two physical
time scales in long GRBs: The first one is associated with the $1$~Hz break and
the second one is associated with the low frequency break due to the finite
burst duration.

%##############################################################################

\section{Dim versus Bright Bursts}

Now we address the following question: Is there any correlation between the 
PDS slope and the burst brightness?

\subsection{PDS Slope Correlates with the Burst Brightness}  

%First we study the longest dim burst individually. 
%We take the four dimmest bursts in the sample with $T_{90}>100$ s.
%Their peak-normalized light curves and PDSs are plotted in Figure~7. 
%The average PDS for the 4 bursts is shown in Figure~7.
%One may notice that the dimmest bursts have PDSs steeper as 
%compared to the brightest bursts shown in Figures 1 and 2.

%%%%%%%%%%%%%%%%%%%%%%%%%%%%%%%%%%%%%%%%%%%%%%%%%%%%%%%%%%%%%%
\smallskip
\begin{figure}
\centerline{\epsfxsize=17cm\epsfysize=17cm {\epsfbox{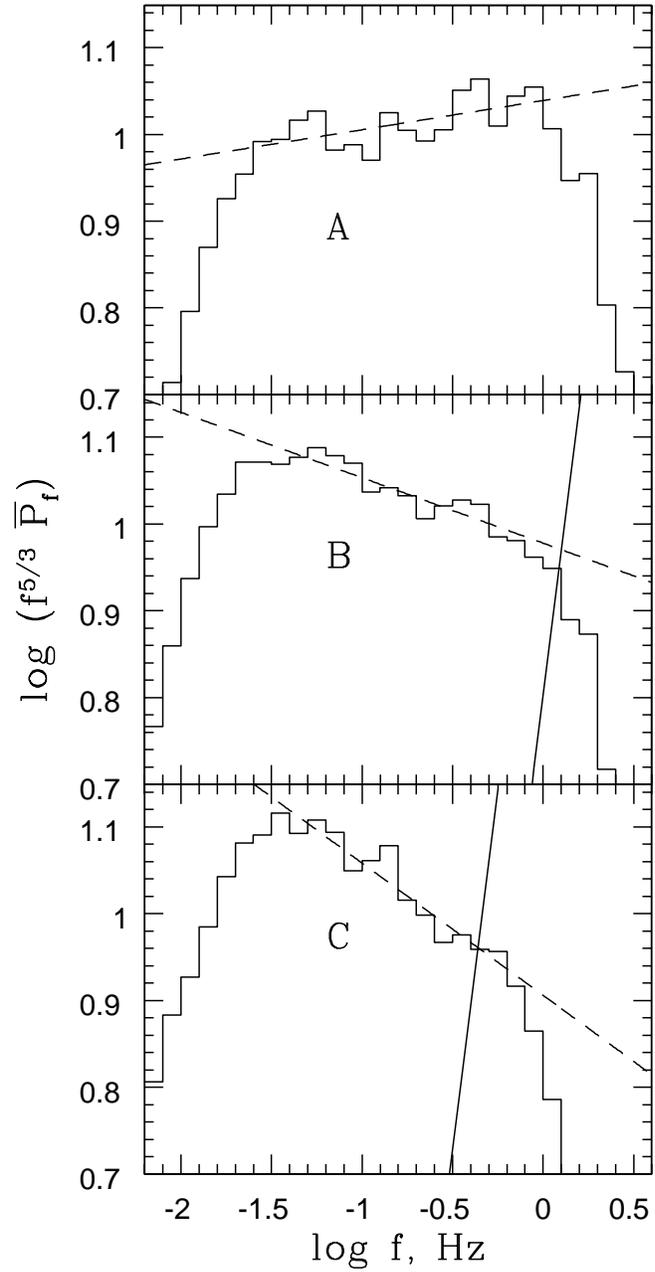}} }
%\bigskip
\caption{ The average PDSs for the three brightness groups A, B, and C.
{\it Dashed lines} show the power-law fits (see Table~3). {\it Solid lines} 
show the Poisson level. 
\label{fig7}}
\end{figure}
%\bigskip
%%%%%%%%%%%%%%%%%%%%%%%%%%%%%%%%%%%%%%%%%%%%%%%%%%%%%%%%%%%%%%

We take the full sample of 527 light curves in channels II+III and divide 
it into 3 groups of different brightnesses: (A) $\Cp>800$, (B) $300<\Cp<800$, 
and (C) $100 < \Cp < 300$. Group A contains 91 bursts, group B 222 bursts, 
and group C 214 bursts. We have calculated the average PDS for each group 
separately employing the peak-normalization. The results are presented in 
Figure~7.
We fitted the average PDSs by power laws, $\log P_f= A+\alpha\log f$.  The  
parameters $A$ and $\alpha$ of the best fits in the range 
$-1.5 < \log f < -0.1$ are listed in Table~3.
We conclude that the average PDS of dim bursts gets markedly steeper.

%*************************************************
%\begin{table}
%\caption[ ]{}
\begin{center}
%{\normalsize
\begin{tabular}{cccc}
            & {\bf Table~3.} &         &        \\ [0.4ex]
\hline\hline
            &                &         &        \\ [0.4ex]
  group     &      A         &    B    &   C    \\ [0.8ex]
  $\alpha$  & $-1.63$        & $-1.74$ & $-1.82$\\ [0.8ex]
  $A$       &   1.04         &   0.98  &   0.91 \\ [0.8ex]
\hline
\end{tabular}
%}
\end{center}
%\end{table}
%*************************************************
\medskip

\subsection{Subtraction of the Poisson Level}

In the PDS calculations we subtracted the Poisson level, which is quite high
for dim bursts (see Fig.~7). 
One therefore should address a technical question: How well is the intrinsic PDS 
restored after subtracting the Poisson level? 

%%%%%%%%%%%%%%%%%%%%%%%%%%%%%%%%%%%%%%%%%%%%%%%%%%%%%%%%%%%%%%
\medskip
\begin{figure}[t]
\centerline{\epsfxsize=9.0cm\epsfysize=9.0cm {\epsfbox{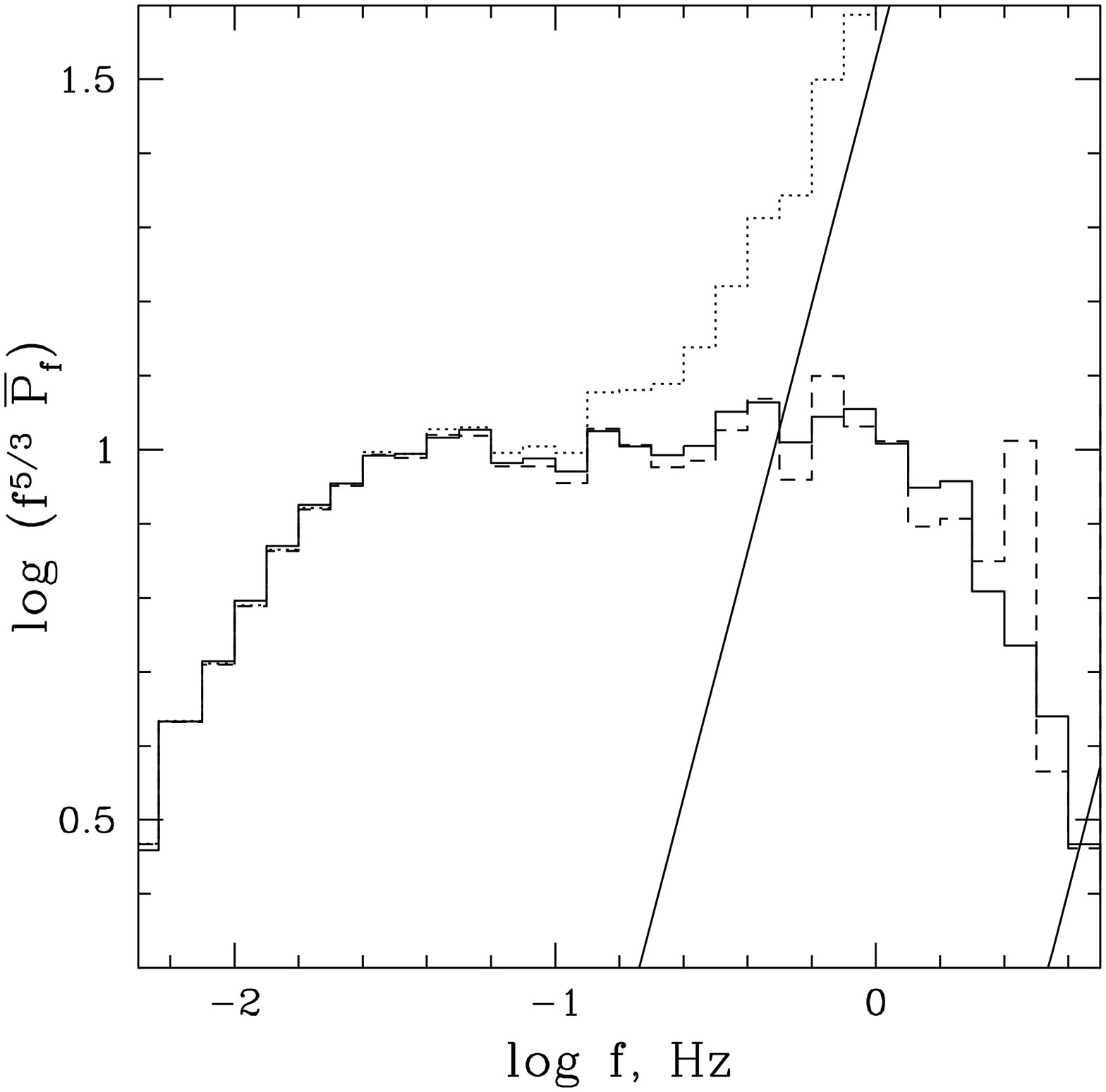}} }
%\bigskip
\caption{ The average PDS of 91 artificial dim bursts created from
group A of the brightest bursts (see Fig.~7). {\it Solid histogram} -- the 
original PDS for group A. {\it Dashed} and {\it dotted histograms}
-- the average PDS of the artificial bursts with and without subtraction of
the Poisson level, respectively. The two solid lines show the average Poisson 
levels for the original and the artificial bursts, respectively. 
\label{fig8}}
\end{figure}
%\bigskip
%%%%%%%%%%%%%%%%%%%%%%%%%%%%%%%%%%%%%%%%%%%%%%%%%%%%%%%%%%%%%%

To investigate this issue we have performed the following test. We take the 
sample of 91 bright bursts (group A) and add a strong Poisson 
noise (with an average level of 25000 counts/bin) to each burst in the group. 
We thus increase artificially the Poisson level by two orders of magnitude.
Alternatively, one may consider this procedure as a rescaling of bright bursts
to a smaller brightness while keeping the Poisson background at the same level. 
Then, we analyze the artificial bursts in the same way as we did with the real 
dim GRBs.
%: (1) set the time-window (see Appendix), 
%(2) determine the Poisson level which is equal to the burst fluence including 
%background in the time-window, (3) subtract the {\it time-averaged} 
%background from the light curve, (4) find the peak, $\Cp$, of the net 
%%%%%%%%%%%%%%%%%%%
%signal\footnote{
Note that even after subtraction of
the time-averaged background, its Poisson noise strongly affects the signal and
may create artificial peaks dominating the true peak of the signal. For the 
artificial bursts, we know the true peak (which is the peak of the original 
bright burst). In real dim bursts we do not know the position of the true peak 
and employ the peak search scheme described in Stern, Poutanen, \& Svensson 
(1999).
%},
%%%%%%%%%%%%%%%%%%%
%(4) calculate the PDS of the signal and subtract the Poisson level, and
%(5) divide the result by $\Cp^2$. Finally, we average the PDS over the sample.

The result is compared with the average PDS of the original sample in 
Figure~8. We find that the subtraction of the Poisson level allows one to 
restore the original PDS, $P_f$, at frequencies where $P_f$ is well below
the Poisson level. Even the 1 Hz break remains present. We conclude that 
the high Poisson level is unlikely to affect significantly the measured PDS 
slope.

%##############################################################################

\section{The Internal Shocks Model}

A likely scenario of GRBs involves internal shocks in a relativistic outflow 
with a Lorentz factor $\Gamma\sim 10^2$ (see, e.g., Piran 1999 for a review). 
The shock develops when an inner faster shell of the outflow tries to overtake 
the 
previous slower shell.
The pulses in a burst are then associated with collisions between the shells. 
%The radius and the time of the collisions are determined
%by initial velocities and times of the shell injection.
The process was simulated numerically by 
Kobayashi, Piran, \& Sari (1997) and Daigne \& Mochkovitch (1998). 

We have performed analogous simulations to see what kind of PDS the model
can produce. The details of the emission mechanism 
are still disputed (see, e.g., M\'esz\'aros \& Rees 1999). For simplicity,
we assume that the energy of shocked plasma is instantaneously radiated.
Then the pressure stays low and its dynamical effect is negligible, i.e., the 
matter is in free motion (coasting) everywhere in the outflow. As time goes on, 
the outflow mass gets concentrated in caustics that
are thin spherically symmetric shells, and the density is reduced between the 
caustics. The caustics have different velocities; they
can catch up with each other and merge with energy release. 

%%%%%%%%%%%%%%%%%%%%%%%%%%%%%%%%%%%%%%%%%%%%%%%%%%%%%%%%%%%%%%
\smallskip
\begin{figure}
\centerline{\epsfxsize=8.4cm {\epsfbox{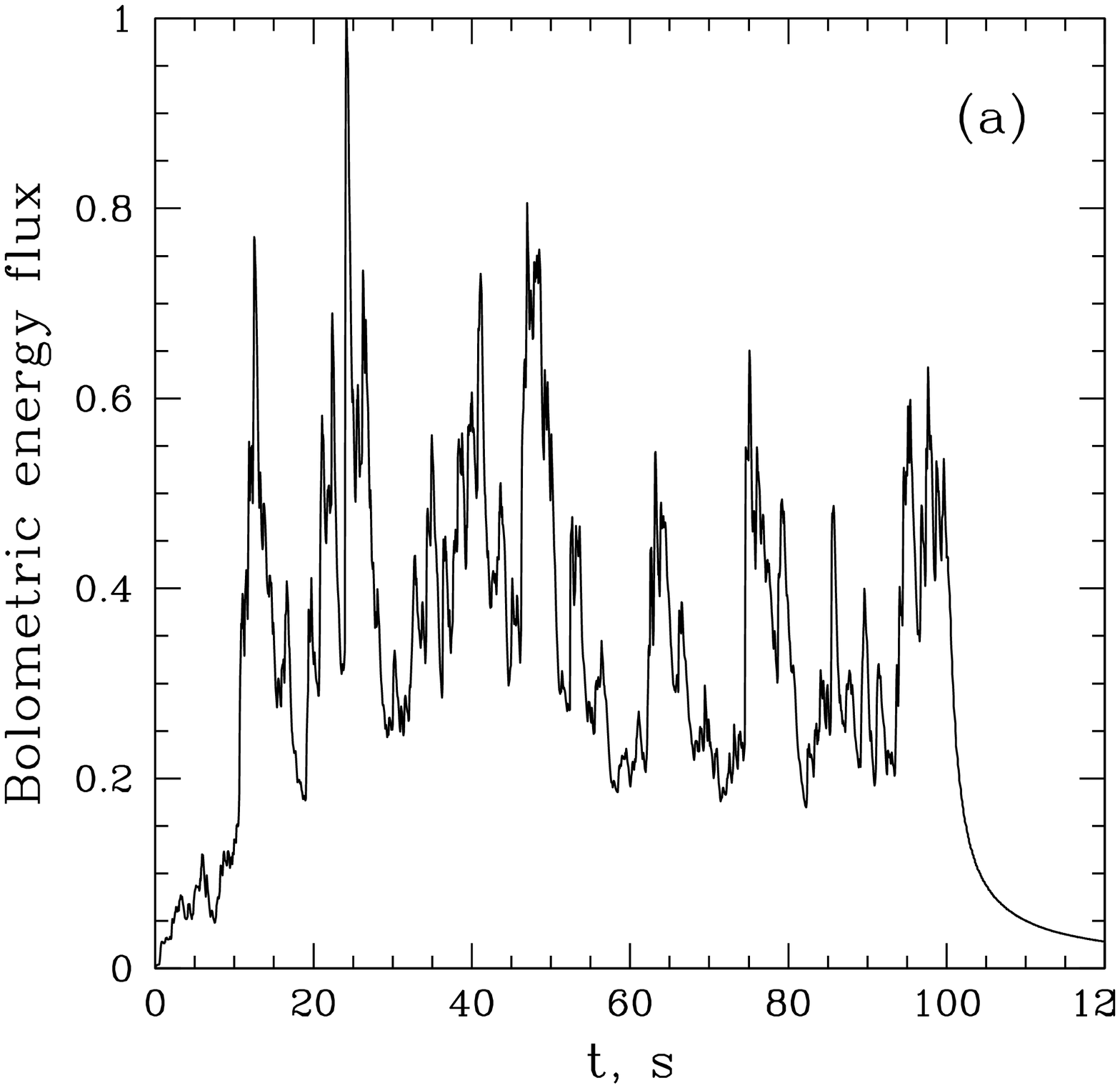}} }
\centerline{\epsfxsize=8.4cm {\epsfbox{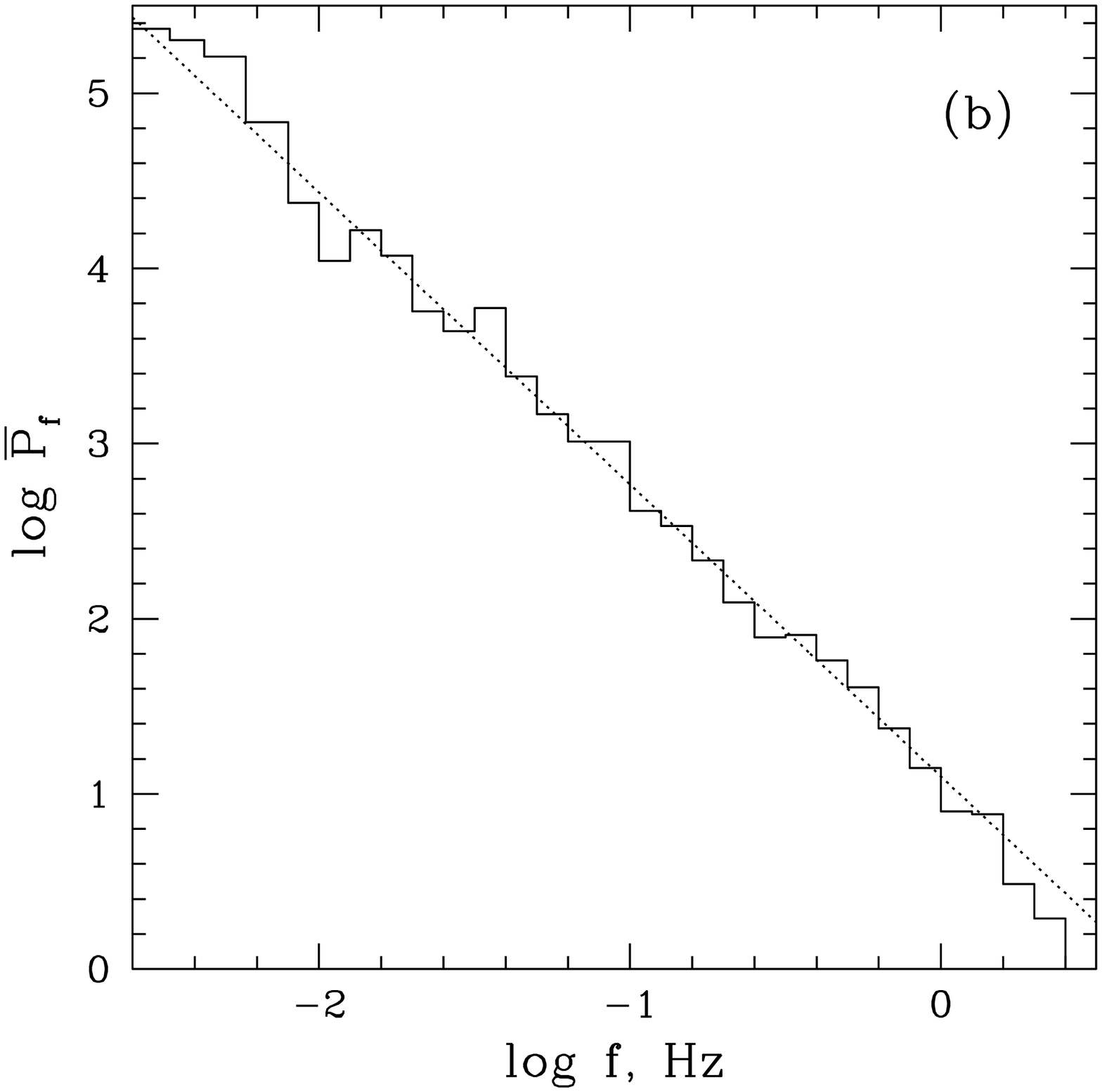}} }
%\bigskip
\caption{ (a) -- One example of the simulated light curves (bolometric energy 
flux
as a function of time). 
%In this model:
%(i) the injection Lorentz factor is randomly chosen between 50 and 150, 
%(ii) the injection density ($m_i$) is a signal with the $-5/3$ PDS and a random 
%phase structure, and (iii) $R_*=10^{14}$~cm.
(b) -- The average PDS of 26 simulated bursts, each is produced by one 
realization
of the random injection process described in the text.
{\it Dotted line} shows the $-5/3$ slope.
\label{fig9}}
\end{figure}
%\bigskip
%%%%%%%%%%%%%%%%%%%%%%%%%%%%%%%%%%%%%%%%%%%%%%%%%%%%%%%%%%%%%%

We take the duration of the outflow, $T_0=\Delta/c=100$~s, where $\Delta$ is the 
distance between the leading and the back fronts of the outflow.
The outflow becomes optically thin when it reaches a radius 
$R_*\sim 10^{13}-10^{14}$~cm $\gg \Delta$. $R_*$ depends on the average density 
and 
Lorentz factor of the outflow; it is taken as a parameter in our simulations. 
We assume that the 
energy released at the optically thick stage is lost as a result of adiabatic 
expansion, and only the radiation produced at $R>R_*$ is observed.
We also neglect the dynamical pressure effect of radiation  
trapped during the optically thick stage. This may be a reasonable 
approximation taking into account 
%the adiabatic cooling and 
the modest
radiative efficiency of internal shocks, which implies that   
the energy density of radiation is small compared to the matter kinetic energy.

The process of energy release is governed by the velocity and the density 
profiles
of the outflow, which 
should be specified as initial conditions at some moment before the outflow 
reaches 
$R_*$. We prepare the initial conditions in the following way.
At a given radius, $R_0\ll R_*$, we ``inject'' the first shell, then (after 
a delay $\delta t$) the second shell and so on, until the last shell. 
It corresponds to ``sampling'' the velocity and density through the outflow when 
it passes radius $R_0$. The initial state is thus represented by the Lorentz 
factors and masses of the discrete shells, $\gamma(t_i)$ and $m(t_i)$, 
$i=1,...,N$.
Here $N$ is the total number of injected shells, $\delta t=t_{i+1}-t_i=T_0/N$ is 
the time scale initially separating the shells.
In the simulations, we take $N=2\times 10^4$. Details of the simulations are 
described
in Beloborodov (1999).

The temporal structure of the injection, $\gamma(t_i)$ and $m(t_i)$, is the main 
issue of the problem. Usually, it was assumed to be a white noise, which leads
to a wrong PDS of the simulated bursts. Trying to reconcile the model with the 
observed $-5/3$ PDS, Panaitescu, Spada, \& M\'esz\'aros (1999) took a correlated
injection with a sinusoidal profile. It leads to extra ``red'' power.
The PDS, however, still does not possess the self-similar structure that we 
observe 
in real GRBs. It appears to be important to have such a structure in the outflow
before it becomes optically thin. We have studied three different regimes of
$\gamma(t_i)$ and $m(t_i)$: (1) white noise, (2) flicker noise, and (3) 
Kolmogorov
noise, and calculated the resulting light curves and their PDSs. 
% The phase structure $\gamma(t_i)$ and $m(t_i)$ is assumed to be Gaussian.  

The best agreement with the observed PDS
is achieved if the initial density and/or the Lorentz factor have a 
Kolmogorov-like
structure.
In Figure 9, we show the result for the case where: (i) the injection Lorentz 
factor 
is randomly chosen between 50 and 150 (white noise), (ii) the injection density, 
$m(t_i)$, is a 
signal with the $-5/3$ PDS and random phase structure (Kolmogorov noise), and 
(iii) $R_*=10^{14}$~cm. 

To the first approximation, the light curve seen by a distant observer goes in 
the 
same order as the injection, i.e., first the observer sees the emission from
internal shocks in the leading part of the outflow and then receives radiation 
from deeper and deeper layers. Different parts of the outflow are causally 
disconnected during the main part of the emission time.
It implies that the light from the back parts cannot overtake  the leading front
of the outflow during the formation of the signal, and radiation from 
the leading and back parts cannot come to the observer in the reverse order.
The observed temporal structure thus conforms to the temporal structure of 
injection, and the resulting slope of the PDS is basically the same as 
prescribed 
to the flow at the inner boundary. For a more detail discussion see Beloborodov 
(1999).

%##############################################################################
\section{Discussion}

\subsection{Relation to the Average Time Profile}

The average time profile (ATP) of GRBs 
%has been studied in detail
% (see, e.g., Norris et al. 1996; Stern 1996).
%The procedure of ATP construction was suggested by Mitrofanov et al. (????):
%the burst light curves are taken peak-normalized, then they are aligned to the 
%peak position, and then averaged. 
%The resulting ATP 
%and it 
was found to follow the stretched exponential of index $1/3$ (Stern 1996). 
Is there any relation between the ATP and the average PDS? One should note two 
important differences between the ATP and the PDS studies:
(i) the $-5/3$ PDS, though affected by the statistical fluctuations, is observed 
in {\it individual} bursts, while the ATP is a purely statistical property of a 
large sample of bursts;
(ii) the ATP is constructed for GRBs of {\it all durations} (and only in this 
case 
it displays the perfect stretched exponential) while the PDS is studied for 
long bursts only. 
Nevertheless, both the ATP and the PDS characterize the stochastic process 
generating GRBs. 
%Moreover, to the first approximation, the average PDS can be 
%considered as a power spectrum of ATP.
%%Both ATP and PDS are related to the auto-correlation function.

%One possible interpretation of the ATP is that it represents a smooth 
%fundamental light curve shape inherent in GRBs and hidden from direct 
%observation in individual GRBs by strong fluctuations superimposed onto the 
%fundamental light curve.
%%pulse (which may appear, e.g., as an "envelope" of the GRB light curves). 
%With this assumption one would expect that the average PDS may also 
%be related to the shape of the fundamental light curve. 
%%If the $-5/3$ law were due to a hidden fundamental shape, 
%One would reveal it only after the averaging, since individual GRBs show 
%strong deviations from any regular behavior. 
%This interpretation is not favored by our findings. We observe the $-5/3$ law
%{\it in individual} bursts. Hence, this law characterizes 
%%the random process generating 
%the fluctuations rather than any smooth light curve.
%We see a signature of a {\it stochastic process} rather than a signature 
%of any fundamental shape of GRB time profiles.

\subsection{The 1 Hz Break}

The sharpness of the break in the average PDS appears to be an important feature 
that constrains models of GRBs. If the signal is produced in the rest frame
of a relativistic outflow then variations of the outflow Lorentz factor, 
$\Gamma$,
from burst to burst would smear out the break. The sharpness of the break then 
implies a narrow dispersion of $\Gamma$, which appears to be unlikely. 

Alternatively, the GRB variability may come from the central engine. 
The break in the PDS might then be related to a typical time scale, $\sim 1$~s, 
in the central engine. 

Finally, one may associate the break with the dynamical time scale corresponding 
to
the inner radius of the optically thin zone of the outflow, $R_*$. Then the
variability 
on time scales shorter than $t_*=R_*/c\Gamma^2$ can be suppressed. In this case, 
however, one should explain why $t_*$ stays the same in different bursts. 

%The break implies that pulses shorter than $\sim$0.5 s are suppressed in the 
%light curve for some physical reason. The deficit of short pulses has also been
%demonstrated directly by analyzing the pulse structure of the light curves 
%(Norris et al. 1996). Note that the sharpness of the break supports the 
%assumption of similar conditions in different GRBs. For example, if the Lorentz
%factor of the emitting gas varied strongly from burst to burst we would observe
%a smooth turnover instead of the sharp break.

\subsection{Dim GRBs}

It has often been hypothesized that dim bursts are at high cosmological 
redshifts. For instance, it is necessarily the case if GRBs have approximately 
the 
same intrinsic luminosity, the so-called "standard candle" hypothesis. 
We can test this hypothesis using the power spectrum analysis.

Suppose that the dim bursts are intrinsically the same as the bright ones.
Then any difference in their average PDS should be due to a cosmological 
redshift. 
First consider the bolometric light curves assuming that their average
PDS follows the $-5/3$ power law. It is easy to see that a redshift, $z$, will 
not change the PDS slope. As we normalize each bursts to its peak before the 
averaging, the effect of a redshift is just stretching the light curve in time,
i.e., precisely the time dilation effect. This will lead to an increase in the 
net normalization of the PDS by a factor of $(1+z)^{1/3}$. The slope does not 
change since the dilation factor $(1+z)$ is the same for each time scale.

One should, however, recall that we observe  
bursts in a limited spectral interval. A redshift then implies a shift of the
signal with respect to our spectral window. E.g., photons detected in 
channel III would originally have been emitted in channel IV.
As we know from Section 6, the 
PDSs are different in different energy channels. Therefore, one expects that 
the PDS slope will change for redshifted GRBs.
%with decreasing brightness of burst.
%of dim burst will differ from that of bright ones.

We have seen in Section 6 that the PDS of bright bursts flattens in the hard 
channels. Hence, a redshift of the bright bursts must be accompanied by a 
flattening of the PDS. Contrary to this behavior, we observe that the PDS of 
dim bursts {\it steepens}. Hence, the evolution of the PDS with brightness is 
inconsistent with the standard candle hypothesis. It implies that the burst 
luminosity function is broad and dim bursts are intrinsically weak. 

Evidence for a broad luminosity function is also found when looking at the 
isotropic luminosities of the bursts with measured redshifts. Note, however,
that the differences in the apparent luminosities could be caused by 
orientation effects if GRBs are beamed. One should not therefore exclude that 
the total intrinsic luminosities are approximately the same. The different
temporal structure of dim bursts may be an important fact in this respect.
%Many of the dim bursts are thus relatively nearby. Their temporal behavior
%differs from the bright bursts. This 
In particular, it suggests that the observed time dilation
of dim bursts (e.g., Norris et al. 1994) may be caused mainly by physical 
processes occurring in the bursts rather than by a cosmological redshift. 
Note that the intrinsic difference of the temporal 
structure of dim GRBs was also found when studying their average time 
profile (see Stern et at. 1997).
The rising part of the ATP does not change with decreasing brightness while 
the decaying part suffers from time dilation. This behavior is inconsistent 
with a cosmological time dilation which should apply equally to both parts 
of the ATP.

\subsection{Internal Shocks}

Further constraints on the internal shock model should be imposed by the 
observed 
dependence of $\alpha$ on photon energy. In particular, the detailed emission 
model studied by Panaitescu, Spada, \& M\'esz\'aros (1999) and Spada, 
Panaitescu,
\& M\'esz\'aros (1999) can be tested.
Besides, the phase structure of the 
signal, neglected so far, should be taken into consideration. In particular, 
the efficiency of the energy release depends on the degree of coherence
of the internal shocks. More detailed analysis in this direction will be given 
elsewhere (Beloborodov 1999).

\acknowledgments

I thank A. F. Illarionov, I. Panchenko, J. Poutanen and R. Svensson for 
discussions. 
This research was supported by the Swedish Natural Science Research Council and 
RFBR 
grant 97-02-16975.
%%We thank A.F. Illarionov, I. Panchenko, and J. Poutanen for discussions. 
%This work was supported by the Swedish Natural Science Research Council,
%the Swedish Royal Academy of Science, the Wennergren Foundation for 
%Scientific Research, a NORDITA Nordic Project grant, and RFFI grant 
97-02-16975.

%\newpage

%\twocolumn

\appendix

\end{document}